\documentclass[twocolumn,5p,amsmath, amssymb, floatfix, sort&compress]{elsarticle}

\usepackage{amsmath}
\usepackage{mathrsfs}
\usepackage{graphicx}
\usepackage{epsfig}
\usepackage{bm}
\usepackage{bbold}
\usepackage{float}
\usepackage{dcolumn}
\usepackage{multirow} 
\usepackage[colorlinks]{hyperref}
\usepackage{color}
\usepackage[dvipsnames]{xcolor}
\usepackage{braket}
\usepackage{dcolumn}
\usepackage{multirow}
\usepackage[titletoc]{appendix}
\usepackage{cleveref}
\usepackage{url}
\usepackage[compat=1.0.0]{tikz-feynman}
\usepackage{feynmf}
\usepackage{lipsum}
\usepackage{soul}

\newcommand{\rbra}[1]{\langle #1||}
\newcommand{\rket}[1]{||#1\rangle}

\renewcommand{\vec}[1]{\boldsymbol{#1}}

\newcolumntype{d}[1]{D{.}{.}{#1}}

\makeatletter
\def\ps@pprintTitle{%
 \let\@oddhead\@empty
 \let\@evenhead\@empty
 \def\@oddfoot{}%
 \let\@evenfoot\@oddfoot}
\makeatother

\begin{document}

\renewcommand{\arraystretch}{1.25}

\title{$\gamma\gamma$ decay as a probe of neutrinoless $\beta\beta$ decay nuclear matrix elements}

\author[dipc,lsc]{B. Romeo\corref{cor1}}
\ead{bromeo@dipc.org}
\author[qpa]{J. Men\'endez}
\ead{menendez@fqa.ub.edu}
\author[lsc,iisb]{C. Pe\~ na Garay}
\ead{cpenya@lsc-canfranc.es}

\address[dipc]{Donostia International Physics Center, 20018 San Sebasti\'an, Spain}
\address[lsc]{Laboratorio Subterr\'aneo de Canfranc,  22880 Canfranc-Estaci\'on, Spain}
\address[qpa]{Department of Quantum Physics and Astrophysics and Institute of Cosmos Sciences, University of Barcelona, 08028 Barcelona, Spain}
\address[iisb]{Institute for Integrative Systems Biology \textup{(I$^2$SysBio)}, Valencia, Spain.}

\cortext[cor1]{Corresponding author}

\begin{abstract}
We study double gamma ($\gamma\gamma$) decay nuclear matrix elements (NMEs) for a wide range of nuclei from titanium to xenon, and explore their relation to neutrinoless double-beta ($0\nu\beta\beta$) NMEs. To favor the comparison, we focus on double-magnetic dipole transitions in the final $\beta\beta$ nuclei, in particular the  $\gamma\gamma$ decay of the double isobaric analog of the initial $\beta\beta$ state into the ground state. For the decay with equal-energy photons, our large-scale nuclear shell model results show a good linear correlation between the $\gamma\gamma$ and $0\nu\beta\beta$ NMEs.  
Our analysis reveals that the correlation holds for $\gamma\gamma$ transitions driven by the spin or orbital angular momentum due to the dominance of zero-coupled nucleon pairs, a feature common to $0\nu\beta\beta$ decay. Our shell-model findings point out the potential of future $\gamma\gamma$ decay measurements to constrain $0\nu\beta\beta$ NMEs, which are key to answer fundamental physics questions based on $0\nu\beta\beta$ experiments.

\end{abstract}

\maketitle

\section{Introduction and main result}
The observation of the decay of an atomic nucleus emitting only two electrons, neutrinoless double-beta ($0\nu\beta\beta$) decay, is the process experimentally most feasible to demonstrate that neutrinos are their own antiparticles~\cite{Avignone08}. Moreover, $0\nu\beta\beta$ decay is one of the most promising probes of physics beyond the standard model (BSM) of particle physics~\cite{Dolinski19}. For instance, the observation of change in lepton number in $0\nu\beta\beta$ decay could help to explain the prevalence of matter in the universe~\cite{Fukugita:1986hr,Davidson08}. Because of this unique potential, a very active program aims to detect $0\nu\beta\beta$ decay. Currently the most stringent constraints reach half-lives longer than $10^{26}$ years~\cite{GERDA20,KamLAND-Zen16,Martin-Albo:2015rhw,CUORE20,Majorana19,CUPID19,EXO19,AMORE19}, and next generation ton-scale experiments are being proposed, among others, for $^{76}$Ge, $^{100}$Mo, $^{130}$Te and $^{136}$Xe nuclei.

Since $0\nu\beta\beta$ decay changes lepton number---no antineutrinos are emitted to balance the two electrons---its decay rate depends on some unknown BSM parameter(s). In the standard scenario that $0\nu\beta\beta$ is triggered by the exchange of known neutrinos, this role is played by a combination of absolute neutrino masses and mixing matrix elements, $m_{\beta\beta}$. The decay rate also depends on a calculable phase-space factor~\cite{Kotila12,Stoica19}, and quadratically on the nuclear matrix element (NME) that involves the initial and final nuclear states~\cite{Engel17}. Thus, $0\nu\beta\beta$ NMEs are needed to interpret current experimental half-life limits and to anticipate the reach of future searches. However, typical NME calculations disagree up to a factor 3~\cite{Bahcall:2004ip,Menendez09,Iwata16,Horoi16b,Coraggio20,Hyvarinen15,Simkovic18,Barea15,Mustonen13,Terasaki20,Rodriguez10,Vaquero14,Song17}, about an order of magnitude on the decay rate. Furthermore, first attempts to obtain more controlled $0\nu\beta\beta$ NMEs using ab initio techniques suggest smaller NME values than most previous studies~\cite{Yao20,Novario21,Belley21}.

A widely explored approach to reduce the uncertainty in $0\nu\beta\beta$ analyses is to study related nuclear observables. Nuclear structure~\cite{Freeman12} or muon capture~\cite{Jokiniemi19b} data are very useful to test nuclear models used to calculate NMEs, but they are not  directly related to $0\nu\beta\beta$ decay. Likewise, $\beta\beta$ decay with neutrino emission shows no apparent correlation with $0\nu\beta\beta$, in spite of both being second-order weak processes sharing initial and final states~\cite{Barabash20}. Useful insights could be gained from nuclear reactions, in the same spirit of the $\beta$ decay information obtained in charge-exchange experiments~\cite{Frekers18,Ejiri19}. Recent efforts include the measurement of nucleon pair transfers~\cite{Brown14,Rebeiro20} and double charge-exchange reactions~\cite{Cappuzzello18}.
The good correlation found between $0\nu\beta\beta$ and double Gamow-Teller transitions~\cite{Shimizu18} could in principle be exploited in double charge-exchange reactions, but the analyses are challenged by tiny cross sections~\cite{Takaki15,Takahisa17} and involved reaction mechanisms~\cite{Lenske19,Santopinto18}.

In this Letter we study the correlation between the NMEs of $0\nu\beta\beta$ and second-order electromagnetic (EM) transitions emitting two photons ($\gamma\gamma$). In fact, the latter were first studied in atoms by Goeppert-Mayer ~\cite{MGoep29,MGoep31}, and it was the extension to the weak interaction which led her to propose the $\beta\beta$ decay~\cite{MGoep35}.
To ensure that nuclear-structure aspects are as similar as possible in the $\gamma\gamma$ and $\beta\beta$ sectors, we focus on EM double-magnetic dipole decays---which depend, like the $0\nu\beta\beta$ operator, on the nuclear spin.
In addition, isospin symmetry assures a good correspondence between the $\gamma\gamma$ and $\beta\beta$ nuclear states if we consider the decay of the double isobaric analogue state (DIAS) of the initial $\beta\beta$ state---an excited state of the final $\beta\beta$ nucleus---into the final $\beta\beta$ state---the ground state (GS) of that nucleus.
Our proposal expands the connections between first-order weak and EM transitions involving isobaric analogue states exploited in the past~\cite{Ejiri68,Fujita11,Ejiri13}.

\begin{figure}[t]
	\centering
	\includegraphics[width=0.49\textwidth]{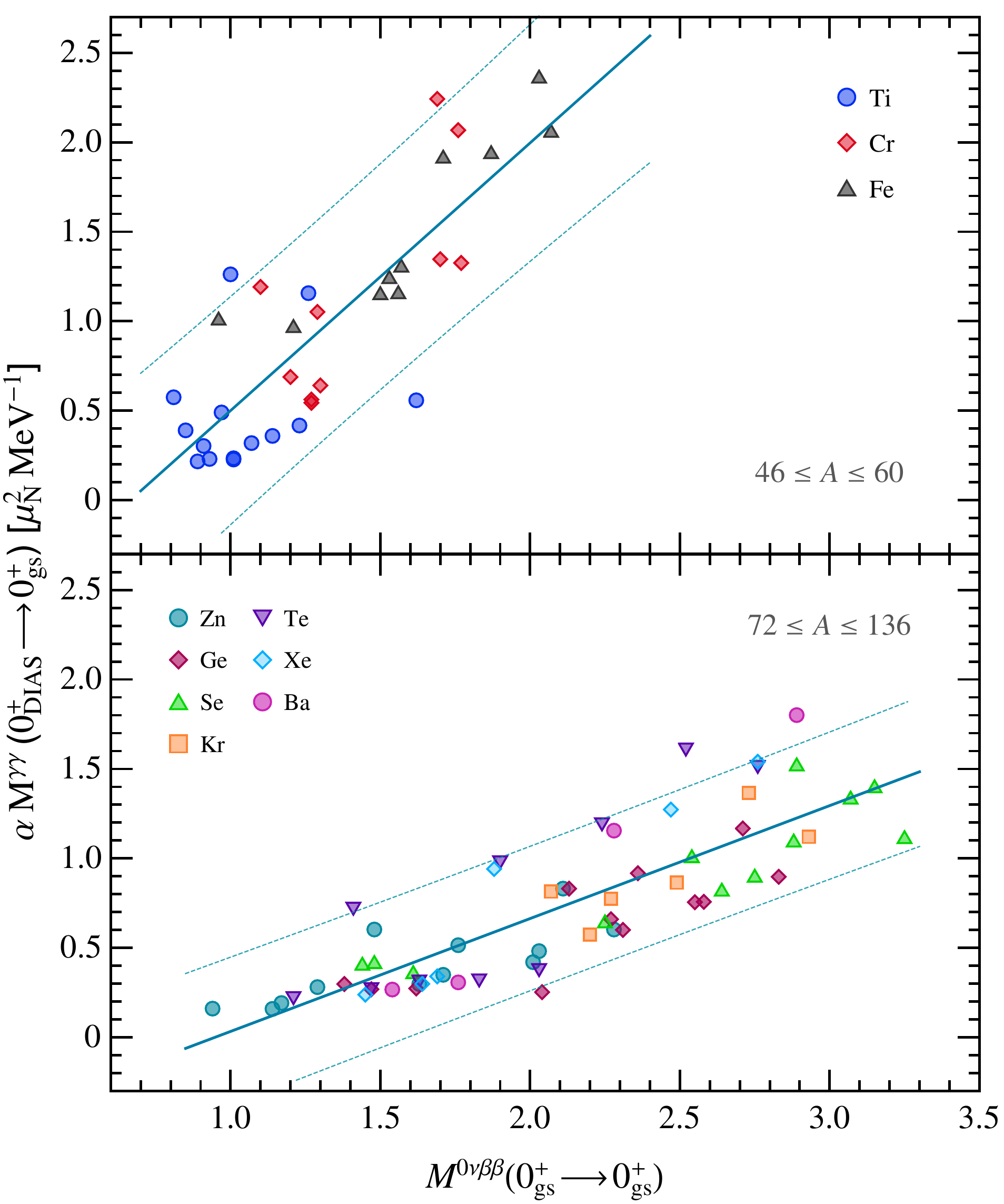}
	\caption{Correlation between $0\nu\beta\beta$ ($M^{0\nu\beta\beta}$, from Ref.~\cite{Shimizu18}) and double-magnetic dipole [$M^{\gamma\gamma}(M1M1)$] NMEs. In the y-axis , $\alpha$ is an isospin factor, see the text.
		Top panel: $^{46-58}\mbox{Ti}$ (blue circles), $^{50-58}\mbox{Cr}$ (red diamonds) and $^{54-60}\mbox{Fe}$ (black triangles). 
		Bottom panel: $^{72-76}\mbox{Zn}$ (light blue circles), $^{74-80}$Ge (violet diamonds), $^{76-82}\mbox{Se}$ (green triangles), $^{82,84}\mbox{Kr}$ (orange squares), 
		$^{124-132}\mbox{Te}$ (violet down triangles), $^{130-134}\mbox{Xe}$ (cyan diamonds) and $^{134,136}\mbox{Ba}$ (magenta circles). 
		The solid line and dashed band represent the best linear fit and prediction band at 90\% confidence level, respectively. }
	\label{fig:correlation}
\end{figure}

Figure~\ref{fig:correlation} summarizes the main result of this work.
We find a good linear correlation between $\gamma\gamma$ and $0\nu\beta\beta$ NMEs obtained with the nuclear shell model, valid across the nuclear chart. 
The upper panel presents results for decays in nineteen nuclei comprising titanium, chromium and iron isotopes with nucleon number $46\leq A\leq 60$. The lower panel covers twenty five nuclei comprising zinc, germanium, selenium, krypton, tellurium, xenon and barium isotopes with $72\leq A\leq 136$.
The correlation is independent on the nuclear interaction used.
Therefore, our findings call for $\gamma\gamma$ calculations with other many-body methods to test to what extent the shell-model correlation in Fig.~\ref{fig:correlation} is universal or depends on the theoretical approach. Indeed the $0\nu\beta\beta$ correlation with double Gamow-Teller matrix elements is common for approaches as different as the nuclear shell model and energy-density functional theory, but it is apparently not fulfilled by the quasiparticle random-phase approximation method.

Second-order EM decays are naturally suppressed with respect to first-order ones. Nevertheless, $\gamma\gamma$ transitions have been measured between $0^+$ first-excited states and GSs, where single-$\gamma$ decay is forbidden~\cite{Watson75,Kramp,Schirmer84}, and, recently, among general nuclear states in competition with $\gamma$ transitions~\cite{Soderstrom20,Walz15}. Future DIAS to GS $\gamma\gamma$ decay measurements, combined with the good linear correlation between NMEs presented in this work, show as a promising tool to give insights on $0\nu\beta\beta$ NMEs. A linear regression analysis supports this potential. Assuming a $\pm15\%$ uncertainty in the branching ratio measurement as in Refs.~\cite{Walz15,Soderstrom20} would lead to a relatively moderate error around $\pm(30\%-40\%)$, dominated by the correlation. This would imply a clear improvement over the large spread in current $M^{0\nu\beta\beta}$ calculations~\cite{Engel17} if the same correlation is found to be valid for other many-body methods used to study $0\nu\beta\beta$ decay.

\section{Electromagnetic DIAS to GS transitions}
The $\gamma\gamma$ decay of a nuclear excited state is an EM process where two photons are emitted simultaneously:
\begin{equation}
\displaystyle  \mathcal{N}_i(p_i)\longrightarrow \mathcal{N}_f(p_f)+\gamma_{\lambda}(k)+\gamma_{\lambda'}(k')\,,
\end{equation}
where $\mathcal{N}_i$, $\mathcal{N}_f$ are the initial and final nuclear states with four-momenta $p_i$ and $p_f$, respectively, and photons have four-momenta $k,k'$ and helicities $\lambda,\lambda'$.

The theoretical framework of nuclear two-photon decay is presented in detail in Refs.~\cite{Friar,FriarRosen,Kramp}. 
The non-relativistic interaction Hamiltonian is given by
\begin{align}
\hat{H}_{I}&=\int{d^4x\,\hat{J}_{\mu}(x)A^{\mu}(x)} \\
&+\frac{1}{2}\int{d^4x\,d^4y\,\hat{B}_{\mu\nu}(x,y)A^{\mu}(x)A^{\nu}(y)}\,, \nonumber
\end{align}
where $A^{\mu}(x)$ denotes the EM field, $\hat{J}_{\mu}(x)$ the nuclear current, and $\hat{B}_{\mu\nu}(x,y)$ is a contact (seagull) operator which represents intermediate nuclear-state excitations not captured by the nuclear model, such as nucleon-antinucleon pairs.  
Perturbation theory up to second order in the photon field leads to the transition amplitudes
\begin{align}
	&\mathcal{M}^{(1)}=\delta(k_0+k'_0+E_f-E_i)  \\
	&\times \sum_n\int d^3\vec{x}\,d^3\vec{y}\,\varepsilon^{*}_{\mu\lambda}(k)\varepsilon^{*}_{\nu\lambda'}(k')\,e^{-i(\vec{k}\cdot\vec{x}+\vec{k}'\cdot\vec{y})}\nonumber \\
	&\times\!\left[\frac{\bra{f}\hat{J}_{\mu}(\vec{x})\ket{n}\bra{n}\hat{J}_{\nu}(\vec{y})\ket{i}}{E_i-k'_0-E_n+i\epsilon}\!+\!\frac{\bra{f}\hat{J}_{\nu}(\vec{y})\ket{n}\bra{n}\hat{J}_{\mu}(\vec{x})\ket{i}}{E_i-k_0-E_n+i\epsilon}\right]\label{eq_M1}, \nonumber\\
	&\mathcal{M}^{(2)}=-(2\pi)\delta(k_0+k_0'+E_f-E_i)  \\
	& \times\! \int d^3\vec{x}\,d^3\vec{y}\,\varepsilon^{*}_{\mu\lambda}(k)\varepsilon^{*}_{\nu\lambda'}(k')e^{-i(\vec{k}\cdot\vec{x}+\vec{k}'\cdot\vec{y})}\bra{f}\hat{B}_{\mu\nu}(\vec{x},\vec{y})\ket{i}, \nonumber
\end{align}
where $\varepsilon_{\mu\lambda}(k)$ is the photon polarization vector.
The initial ($\ket{i}$), intermediate ($\ket{n}$) and final ($\ket{f}$) nuclear states have energies $E_i$, $E_n$ and $E_f$, respectively.
The amplitude $\mathcal{M}^{(2)}$ can be neglected for DIAS to GS transitions, in the absence of subleading two-nucleon currents, because it involves a one-nucleon operator in isospin space~\cite{Kramp}.

It is very useful to perform a multipole decomposition of the $\gamma\gamma$ amplitude, because nuclear states have good angular momentum. The expansion involves electric ($E$) and magnetic ($M$) multipole operators with angular momentum $L$, denoted as $X$. The transition amplitude sums over multipoles, which factorize into a geometrical (phase space) factor and the generalized nuclear polarizability, $\mathcal{P}_J$, containing all the information on the nuclear structure and dynamics~\cite{Kramp}:
\begin{align}
\label{GeP1}
&\mathcal{P}_J(X'X;k_0,k'_0)=2\pi(-1)^{J_f+J_i}\!\sqrt{(2L+1)(2L'+1)}  \\ 
&\times \sum_{n,J_n}\Biggl[
\begin{Bmatrix}
L & L' & J\\
J_i & J_f & J_n
\end{Bmatrix} 
\frac{\rbra{J_f}\widetilde{\mathcal{O}}(X)\rket{J_n}\rbra{J_n}\widetilde{\mathcal{O}}(X')\rket{J_i}}{E_n-E_i+k'_0} \Biggr.\nonumber \\
&+(-1)^Y
\begin{Bmatrix}
L' & L & J\\
J_i & J_f & J_n
\end{Bmatrix}  
\left.\frac{\rbra{J_f}\widetilde{\mathcal{O}}(X')\rket{J_n}\rbra{J_n}\widetilde{\mathcal{O}}(X)\rket{J_i}}{E_n-E_i+k_0}\right], \nonumber
\end{align}
where the $6j$-symbols depend on the total angular momenta of the initial, intermediate, and final states
$J_{i}$, $J_n$, $J_f$ and $Y=J-L-L'$.
The reduced matrix elements of the EM multipole operators involve the photon energy: $\widetilde{\mathcal{O}}(X)\propto k_0^L$, as well as the nucleon radial $r$, angular spherical harmonics $Y_L$, orbital angular momentum $\bm{l}$, and spin $\bm{s}$ operators. 

Double EM and weak decays involve different nuclei:
$^{A}_{Z}Y^*_{N}\!\rightarrow^{A}_{Z}\!Y_{N}+2\gamma$ vs $^{A}_{Z-2}X_{N+2}\!\rightarrow^{A}_{Z}\!Y_{N}+2e^-$, with $N,Z$ the neutron and proton number.
In order to study the correlation between ${0\nu\beta\beta}$ and $\gamma\gamma$ NMEs,
we focus on the $\gamma\gamma$ decay of the DIAS of the initial $\beta\beta$ state. This is an excited state, with isospin $T=T_z+2$, of the $\beta\beta$ daughter nucleus with isospin third component $T_z=(N-Z)/2$. The DIAS $\gamma\gamma$ decay to the GS---the final $\beta\beta$ state---with $T=T_z$, thus connects states with the same isospin structure as $\beta\beta$ decay: an initial state with isospin $T_i=T_f+2$ with a final one with $T_f=T_z$.
Since isospin symmetry holds very well in nuclei, we expect the nuclear structure aspects of DIAS to GS $\gamma\gamma$ and $0\nu\beta\beta$ transitions to be very similar.
Altogether, the $\gamma\gamma$ decay involves the following positive-parity $J_i=J_f=0$ nuclear states:
\begin{align}
\label{eq:0DIAS}
\ket{{0}_i^+}_{\gamma\gamma}&\equiv\ket{0^+_{i}}_{\beta\beta}\!(\text{DIAS})=
\frac{T^{-}T^{-}}{K^{1/2}}\ket{0^+_{i}}_{\beta\beta}\,,\\
\ket{0^+_f}_{\gamma\gamma}&\equiv\ket{0^+_{f}}_{\beta\beta}\,,
\end{align}
with ${K}$ a normalization constant and $T^{-}=\sum_{i}^{A}t^{-}_i$ the nucleus isospin lowering operator, which only changes $T_z$.

Angular momentum and parity conservation impose that transitions between $0^+$ states just involve the zero-multipole polarizability $\mathcal{P}_0$, with two $EL$ or $ML$ operators. In the long wave approximation $k_0|\vec{x}|\ll 1$, satisfied when 
$Q=E_i-E_f\sim1-10$~MeV, dipole ($L=1$) decays dominate.
Since the nuclear spin is key for $0\nu\beta\beta$ decay, we focus on double-magnetic dipole ($M1M1$) processes, governed by the operator
\begin{equation}
\vec{M1}=\mu_N\sqrt{\frac{3}{4\pi}}\sum_{i=1}^A \left(g^l_{i}\vec{l}_i+g^s_{i}\vec{s}_i\right)\,,
\end{equation}
with $\mu_N$ the nuclear magneton, and the neutron ($n$) and proton ($p$) spin and orbital $g$-factors: $g^s_n=-3.826$, $g^s_p=5.586$, $g^l_n=0$, $g^l_p=1$.

For M1M1 transitions Eq.~\eqref{GeP1} factorizes and the expression can be written in terms of a single NME $\mathcal{M}^{\gamma\gamma}(M1M1,\Delta\varepsilon)$:
\begin{align}
&\mathcal{P}_0(M1M1,k_0,k_0')={\frac{2\pi}{3\sqrt{3}}}\,k_0k'_0\,\mathcal{M}^{\gamma\gamma}(M1M1,\Delta\varepsilon)\,, \\
&\mathcal{M}^{\gamma\gamma}(M1M1,\Delta\varepsilon)=\sum_{n}\frac{\rbra{0^+_f}\bm{M1}\rket{1^+_n}\rbra{1^+_n}\bm{M1}\rket{0^+_i}}{\varepsilon_n\left(1-\frac{\Delta\varepsilon^2}{2\varepsilon_n^2}\right)}\,,
\label{eq:P01}
\end{align}
which depends on the energy difference between the two photons, $\Delta\varepsilon=k_0-k_0'$, and on $\varepsilon_n=E_n-(E_i+E_f)/2$. The $\vec{M1}$ operator demands $1^+$ intermediate states. In order to avoid the dependence on the photon energies, which depends on the nucleus, we require that the two photons share the transition energy: $k_0=k'_0=Q/2$ and thus $\Delta\varepsilon=0$. In this case we can define the following nuclear matrix element $M^{\gamma\gamma}(M1M1)=\mathcal{M}^{\gamma\gamma}(M1M1,0)$:
\begin{equation}
M^{\gamma\gamma}(M1M1)=\sum_{n}\frac{\rbra{0^+_f}\bm{M1}\rket{1^+_n}\rbra{1^+_n}\bm{M1}\rket{0^+_i}}{\varepsilon_n}\,.
\label{eq:M1M1}
\end{equation}
In the following, we calculate the NMEs in Eq.~\eqref{eq:M1M1}. Note that the $\gamma\gamma$ measurement to constrain these NMEs requires an experimental setup for $k_0\simeq k_0'$. This is needed because in transitions between the DIAS and the ground state $Q$ and hence $\Delta\varepsilon$ can easily be of the order of $10$~MeV, exceeding the value of $\varepsilon_n$.

\section{Nuclear shell model calculations} We calculate $M^{\gamma\gamma}(M1M1)$ for a broad range of $46\leq A \leq 136$ nuclei in the framework of the nuclear shell model~\cite{MPinedo,Otsuka20,Brown01}.
We cover three different configuration spaces spanning the following harmonic oscillator single-particle orbitals for protons and neutrons: 
i) $0f_{7/2}$, $1p_{3/2}$, $0f_{5/2}$ and $1p_{1/2}$ (\textit{pf} shell) with the KB3G~\cite{PovesKB3G} and GXPF1B~\cite{HonmaGXPF} effective interactions;
ii) $1p_{3/2}$, $0f_{5/2}$, $1p_{1/2}$ and $0g_{9/2}$ (\textit{pfg} space)  with the GCN2850~\cite{Caurier08}, JUN45~\cite{HonmaJUN45} and JJ4BB~\cite{BrownJJ4BB} interactions; and
iii) $1d_{5/2}$, $0g_{7/2}$, $2s_{1/2}$, $1d_{3/2}$ and $0h_{11/2}$ (\textit{sdgh} space) with the GCN5082~\cite{Caurier08} and QX~\cite{QiQX} interactions.
All the interactions are isospin symmetric.
For our calculations we use the shell model codes ANTOINE~\cite{FNowacki,MPinedo} and NATHAN~\cite{MPinedo}. The $0\nu\beta\beta$ NMEs, calculated with the same configuration spaces and nuclear interactions, are taken from Ref.~\cite{Shimizu18}.

First, we calculate the final $\gamma\gamma$ state and the initial $\beta\beta$ one, 
which we rotate in isospin to obtain its DIAS as in Eq.~\eqref{eq:0DIAS}.
Next, we build a finite set of intermediate states $\{1_n^{+}\}$ with the Lanczos strength function method, taking as doorway state the isospin $T_n=T_f+1$ projection of the isovector $\vec{M1}$ operator applied to the final state: $P_{T=T_z+1}\,\vec{M1}_{\text{IV}}\ket{0_f^+}$.
This guarantees intermediate states with correct angular momentum and isospin.

\begin{figure}[t]
	\centering
	\includegraphics[width=0.48\textwidth]{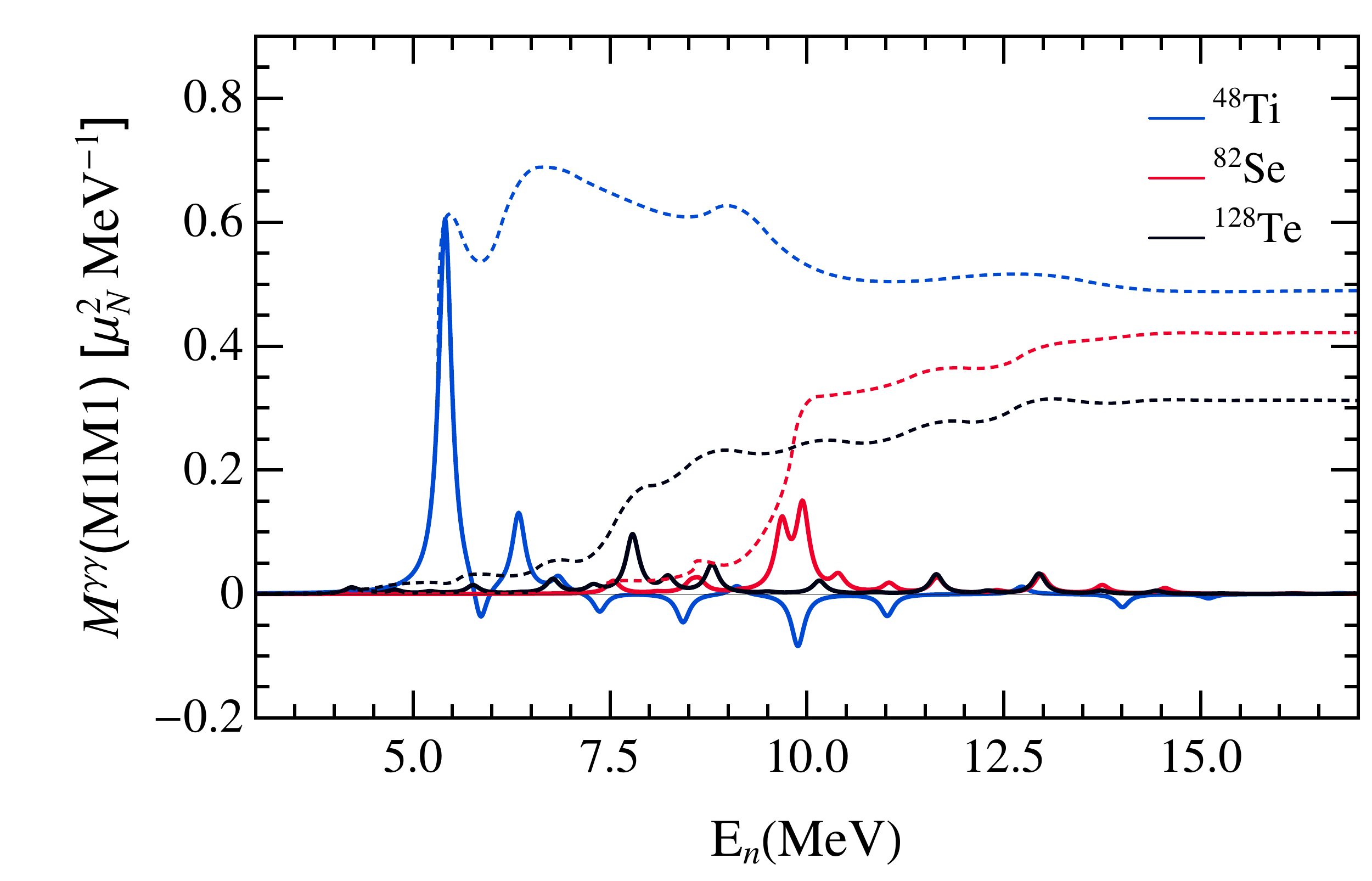}
	\caption{Contribution (solid lines) and cumulative (dashed lines) values of the $M^{\gamma\gamma}(M1M1)$ NME as a function of the excitation energy of the intermediate states $E_n$.
		The results, for the $0_{\text{DIAS}}^+\rightarrow 0^+_{\text{gs}}$ transition in $^{48}$Ti, $^{82}$Se and $^{128}$Te, are smoothed with a Lorentzian of width 0.1 MeV.}
	\label{fig:convergence}
\end{figure}

We evaluate the energy denominator $\varepsilon_n$ using experimental energies when possible~\cite{NucDataServ,Wang_2017}.
For $^{48}$Ti, $E_f$, $E_i\text{(DIAS)}$ and also the energy of a $T=T_z+1$ state (${6^+}$) are known. We use the latter, together with the calculated energy difference between the $6^+$ and $1^+$ states with $T=T_z+1$, to fix the energy of the intermediate states $E_n$.
With this experimental input,
$M^{\gamma\gamma}(M1M1)$ only varies the result obtained with calculated energies by $0.2\%$.
Isospin-breaking effects cancel in $\varepsilon_n$ to a very good approximation~\cite{Roca-Maza19}.
Therefore, in nuclei with unknown energy of the DIAS or $T=T_z+1$ states, we use experimental data on states of the same isospin multiplet in neighboring nuclei: the $\beta\beta$ parent to fix $E_i$, and the $\beta\beta$ intermediate nucleus---when available---for $E_1$.
Using these experimental energies modifies $M^{\gamma\gamma}(M1M1)$ results by less than 5\%.

\section{Results}
With these ingredients we evaluate Eq.~\eqref{eq:M1M1}.
Figure~\ref{fig:convergence} shows $M^{\gamma\gamma}(M1M1)$ as a function of the excitation energy of the intermediate states, for nuclei covering the three configuration spaces: $^{48}$Ti, $^{82}$Se and $^{128}$Te. The Lanczos strength function gives converged results to $\sim1\%$ after $50-100$ iterations.
Figure~\ref{fig:convergence} illustrates that, in general, intermediate states up to $\sim15$~MeV can contribute to the double-magnetic dipole NME, and that only a few states dominate each transition.
The comparison between weak and EM decays needs to take into account
that while $0\nu\beta\beta$ changes $N$ and $Z$ by two units, they are conserved in $\gamma\gamma$ decay. This is achieved by comparing isospin-reduced NMEs or, alternatively, by
including the ratio of Clebsch-Gordan coefficients dictated by the 
Wigner-Eckart theorem~\cite{Edmonds}: $\alpha=\sqrt{\frac{3}{2}}
C^{T_f,2,T_f+2}_{T_f,2,T_f+2}/C^{T_f,2,T_f+2}_{T_f,0,\quad T_f}=\frac{1}{2}\sqrt{(2+T_f)(3+2T_f)}$.

Figure~\ref{fig:correlation} shows the good linear correlation between $0\nu\beta\beta$ NMEs and double-magnetic dipole NMEs obtained with bare spin and orbital $g$-factors.
We observe essentially the same correlation when using effective $g$-factors that give slightly better agreement with experimental magnetic dipole moments and transitions: $g^s_i(\text{eff})=0.9 g^s_i$, $g^l_{p}(\text{eff})=g^l_p+ 0.1$, $g^l_{n}(\text{eff})=g^l_n-0.1$  in the \textit{pf} shell~\cite{Honma04}; and $g^s_i(\text{eff})=0.7 g^s_i$ for \textit{pfg} nuclei~\cite{Honma09}.

We have performed a linear regression analysis to the data leading to the correlation in Fig.~\ref{fig:correlation}. A fit to the function $M^{0\nu\beta\beta}=a+bM^{\gamma\gamma}$ gives best-fit parameters $a=0.872, b=0.459$ for $46\leq A\leq60$ (top panel), and $a=1.29,b=1.11$ for $72\leq A\leq136$ (bottom). The best fit is shown with a solid line, while prediction bands at 90\% confidence level (CL) are given in dashed lines. The correlation coefficients for the top and bottom panels are $\rho=0.83$ and $\rho=0.84$, respectively. The 90\% CL bands could be combined with a hypothetical measurement of the $\gamma\gamma$ M1M1 decay to obtain a $M^{0\nu\beta\beta}$ NME. A branching ratio measurement with a $\pm15\%$ uncertainty~\cite{Walz15,Soderstrom20} combined with the linear correlation would lead to a relatively moderate error around $\pm(30\%-40\%)$.

\begin{figure}[t]
	\centering
	\includegraphics[width=0.48\textwidth]{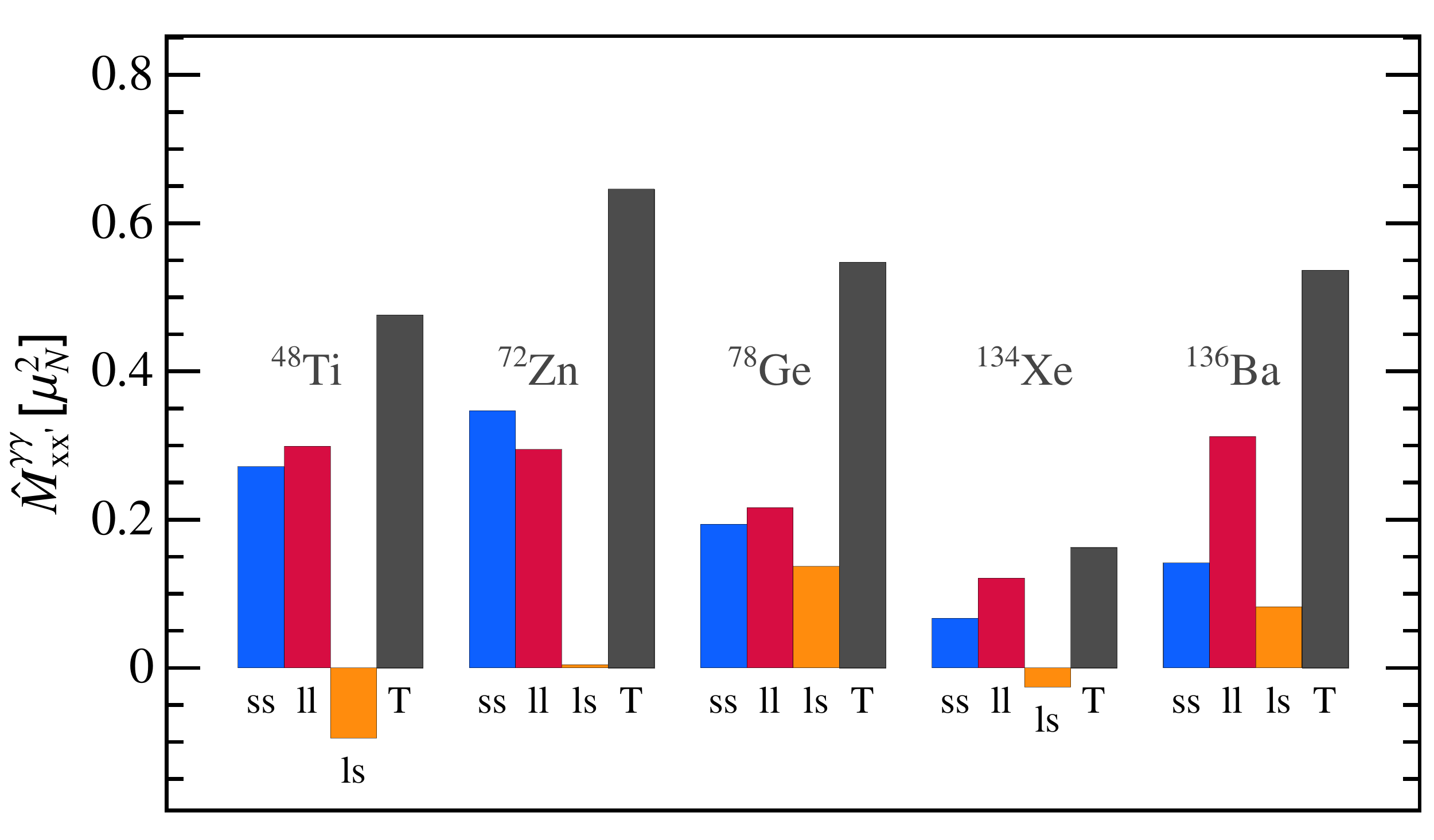}
	\caption{Different contributions to the numerator NME $\hat{M}^{\gamma\gamma}$ for several nuclei: total (T), spin $\hat{M}^{\gamma\gamma}_{ss}$ (ss), orbital $\hat{M}^{\gamma\gamma}_{ll}$ (ll) and interference $\hat{M}^{\gamma\gamma}_{ls}$ (ls) terms.}
	\label{fig:M1M1components}
\end{figure}

The slope of the linear correlation between $\gamma\gamma$ and $0\nu\beta\beta$ NMEs in Fig.~\ref{fig:correlation} only depends mildly on the mass number, being slightly larger in the \textit{pf} shell than for \textit{pfg} and \textit{sdgh} nuclei.
This distinct behaviour is due to the energy denominator in $M^{\gamma\gamma}(M1M1)$: when only the numerator in Eq.~\eqref{eq:P01} is considered, $\hat{M}^{\gamma\gamma}$, the same linear correlation is common to all nuclei. Ultimately, this mild dependence on the energy denominator is key for the good correlation between $M^{\gamma\gamma}$ and $M^{0\nu\beta\beta}$.

The small dependence on the energy denominator is illustrated by Fig.~\ref{fig:convergence}: 
the intermediate states that contribute more to $M^{\gamma\gamma}(M1M1)$ lie systematically at lower energies in \textit{pf}-shell nuclei, compared to $A\geq 72$ systems.
In fact, the ratio of average energy of the dominant states contributing to $M^{\gamma\gamma}(M1M1)$ in the \textit{pf} shell over the \textit{pfg}$-$\textit{sdgh} spaces matches very well  
the ratio of the slopes in the top and bottom panels of Fig.~\ref{fig:correlation}.
Also, in the bottom panel of Fig.~\ref{fig:correlation} heavier nuclei (Te, Xe, Ba) calculated with the GCN5082 interaction appear in the upper part of the correlation band. This is partly due to a mildly smaller energy denominator and also because of a slightly larger contribution of the orbital angular momentum component of the $M1M1$ operator.

We can gain additional insights on the $\gamma\gamma-0\nu\beta\beta$ correlation by decomposing the double-magnetic dipole NME into spin, orbital and interference parts.
Since the energy denominator plays a relatively minor role, we focus on the changes in the numerator matrix element:
$\hat{M}^{\gamma\gamma}=\hat{M}^{\gamma\gamma}_{ss}+\hat{M}^{\gamma\gamma}_{ll}+\hat{M}^{\gamma\gamma}_{ls}$.
Figure \ref{fig:M1M1components} shows the decomposition for the $\gamma\gamma$ decay of several nuclei.
In some cases like $^{72}$Zn, the spin part dominates.
Here, since $\hat{M}^{\gamma\gamma}_{ss}$ is proportional to the double Gamow-Teller operator, a very good correlation with $0\nu\beta\beta$ is expected~\cite{Shimizu18}.
In contrast, the orbital $\hat{M}^{\gamma\gamma}_{ll}$ part dominates in $^{134}$Xe or $^{136}$Ba, $sdgh$ nuclei with an $l=5$ orbital.
Remarkably, these nuclei follow the common trend in Fig.~\ref{fig:correlation}, which means that the correlation with $0\nu\beta\beta$ decay is not limited to operators driven by the nuclear spin.
The interference $\hat{M}^{\gamma\gamma}_{ls}$ is generally smaller, and can be of different sign to the dominant terms. In fact, Fig.~\ref{fig:M1M1components} also shows that the spin and orbital contributions to $\gamma\gamma$ decay always have the same sign, preventing a cancellation that would blur the correlation with $0\nu\beta\beta$ decay.

\begin{figure}[t]
	\centering
	\includegraphics[width=0.49\textwidth]{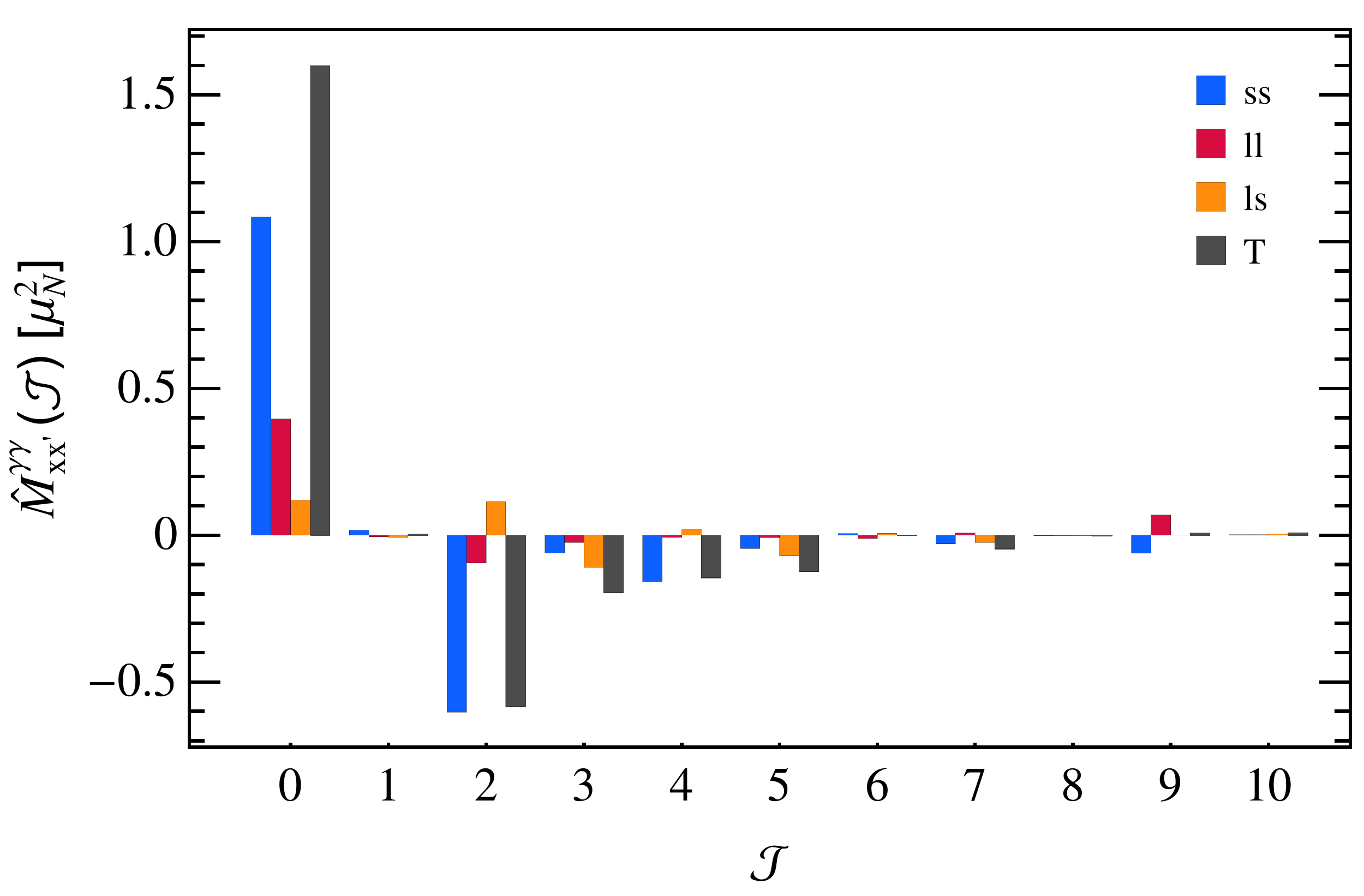}
	\caption{Decomposition of the $^{136}\mbox{Ba}$ numerator NME $\hat{M}^{\gamma\gamma}$, in terms of the two-nucleon angular momenta $\mathcal{J}$: total (T), spin $\hat{M}^{\gamma\gamma}_{ss}$ (ss), orbital $\hat{M}^{\gamma\gamma}_{ll}$ (ll) and interference $\hat{M}^{\gamma\gamma}_{ls}$ (ls) parts.}
	\label{fig:M1M1Jdecomposition}
\end{figure}

Figure~\ref{fig:M1M1Jdecomposition} investigates further the relation between spin and orbital $\gamma\gamma$ contributions, decomposing the NMEs in terms of the two-body angular momenta $\mathcal{J}$ of the two nucleons involved in the transition.
Analogously to $0\nu\beta\beta$ NMEs~\cite{Iwata16,Menendez09}, $\hat{M}^{\gamma\gamma}$ is dominated by the contribution of $\mathcal{J}=0$ pairs, partially canceled by that of $\mathcal{J}>0$ ones.
This behaviour is common to $\hat{M}^{\gamma\gamma}_{ss}$ and $\hat{M}^{\gamma\gamma}_{ll}$, with a more marked cancellation in the spin part, as expected due to the spin-isospin SU(4) symmetry of the isovector spin operator~\cite{Menendez16,Simkovic18}.
The $\mathcal{J}=0$ dominance suggests that spin and orbital $\mathcal{S}=\mathcal{L}=0$ pairs are the most relevant in $\gamma\gamma$ DIAS to GS transitions, implying that $s_1 s_2=(\mathcal{S}^2-3/2)/2<0$, and likewise $l_1 l_2<0$. Since the spin and orbital isovector $g$-factors also share sign, the hierarchy in Fig.~\ref{fig:M1M1Jdecomposition} explains the absence of cancellations that leads to the $\gamma\gamma$ correlation with $0\nu\beta\beta$ decay.

The shell-model $0\nu\beta\beta$ nuclear matrix elements in Fig.~\ref{fig:correlation} have been obtained with axial coupling $g_A=1.27$. While the nuclear shell model is known to overestimate $\beta$~\cite{Martinez-Pinedo96} and two-neutrino $\beta\beta$~\cite{Caurier12,Neacsu15,Horoi13} matrix elements---a feature usually known as ``$g_A$ quenching''---the need and amount of ``quenching'' required by shell-model $0\nu\beta\beta$ NMEs is uncertain. For instance, the larger $0\nu\beta\beta$-decay momentum transfer may imply different sensitivity to missing nuclear correlations and two-body currents, the main aspects that cause the deficiencies in $\beta$ and two-neutrino $\beta\beta$ calculations~\cite{Engel17,Gysbers19}---two-body currents are expected to be less relevant for $0\nu\beta\beta$ NMEs~\cite{Menendez11,Wang18}. In contrast, ab initio calculations in light~\cite{Pastore17,King20,Gysbers19} and middle-mass nuclei~\cite{Gysbers19,Stroberg21} describe well $\beta$-decay matrix elements without additional adjustments. A comparison to the first ab initio $0\nu\beta\beta$ NMEs~\cite{Yao20,Belley21,Novario21} and also to a recently proposed hybrid approach that combines ab initio short-range correlations with the nuclear shell model~\cite{Weiss21} suggests that $0\nu\beta\beta$ NMEs in the shell model may be moderately overestimated---by several tens of percent---in a relatively similar way for all $\beta\beta$ emitters. This would imply a similar correlation to the one presented in Fig.~\ref{fig:correlation} but with $a$ and $b$ parameters modified according to the possible overestimation of the shell-model $0\nu\beta\beta$ NMEs.

Future work includes evaluating two-nucleon current contributions to double-magnetic dipole~\cite{Bacca14} and $0\nu\beta\beta$~\cite{Menendez11,Gysbers19} decays, but we do not expect these corrections to alter significantly the NME correlation. In contrast, the recently-proposed leading-order contact contribution could modify sizeably $0\nu\beta\beta$ NMEs~\cite{Cirigliano18,Wirth21,Jokiniemi21}, but note that short-range NMEs can also be correlated to the $0\nu\beta\beta$ ones in Fig.~\ref{fig:correlation}~\cite{Menendez18}. In addition, the correlation observed here can be tested with other many-body approaches such as energy-density functional theory~\cite{Robledo18,Rodriguez10,Vaquero14}, the interacting boson model~\cite{Barea15,Otsuka78}, the quasiparticle ramdom-phase approximation (QRPA)~\cite{Hyvarinen15,Delion03} or ab initio methods~\cite{Yao20,Stroberg19,Novario21,Belley21}.

\section{Summary}\label{summary}
We have observed a good linear correlation between $0\nu\beta\beta$ NMEs and $\gamma\gamma$ ones when the two photons share the energy of the decay. For our shell model calculations, the correlation holds across the nuclear chart, independently on the nuclear interaction used. While the correlation should also be tested with other leading many-body methods used to study $0\nu\beta\beta$ decay, this suggests a new avenue to reduce $0\nu\beta\beta$ NME uncertainties if double-magnetic dipole DIAS to GS $\gamma\gamma$ transitions can be measured, especially on the most relevant $0\nu\beta\beta$ nuclei.
In fact, first steps in this direction are underway: Valiente-Dob\'on {\it et al.}~\cite{Valiente-Dobon} recently proposed a flagship experiment to determine the conditions of a future program to measure the $\gamma\gamma$ decay of the $^{48}$Ca DIAS in $^{48}$Ti.
Even though these experiments are challenging due to the competition with single-$\gamma$, $E1E1$ $\gamma\gamma$ and nucleon-emission channels, their potential should not be underestimated. Next generation $0\nu\beta\beta$ experiments imply a significant investment with the promise to fully cover the inverted neutrino-mass hierarchy region~\cite{Murayama:2003ci}, but current NME uncertainties may limit the reach of the proposals under discussion.

\section*{Acknowledgement}
We thank J.~J.~Valiente-Dob\'on for illuminating discussions. B.~R. warmly acknowledges support from the NEXT Collaboration.
This work was supported in part by the ``Ram\'on y Cajal'' program with grant RYC-2017-22781, and grants CEX2019-000918-M, PID2020-118758GB-I00 and RTI2018-095979-B-C41 funded by MCIN/AEI/10.13039/501100011033 and, as appropriate, by "ESF Investing in your future".

\bibliographystyle{apsrev4-1}

\end{document}